\documentstyle{article}
\begin{document}
\begin{center}
\Large\textbf{ Reissner-Nordstrom black hole in noncommutative
spaces.}
\end{center}

\begin{center}
\textbf{S. A. Alavi}

\textit{Department of Physics,  Sabzevar Tarbiat Moallem university,  P. O. Box 397, Sabzevar, 
 Iran}\\

\textit{}\\

\textit{Email: alavi@sttu.ac.ir}
 \end{center}

\emph{ We investigate the behaviour of a non-commutative radiating
Reissner-Nordstrom(Re-No)black hole. We find  some interesting
results : a). the existence of a minimal non-zero mass to which
 the black hole can shrink. b). a finite maximum temperature that the
black hole can reach before cooling down to absolute zero. c)
compared to the neutral black holes  the effect of
charge is to increase  the minimal non-zero mass and  lower the
maximum temperature. d) the absence of any curvature singularity.
We also derive some essential thermodynamic quantities from which
we study the stability of the black hole. Finally we  find an upper bound for the non-commutativity parameter $\theta$.}\\



 \section{Introduction}
 
It is generally believed that the picture of  continuous space-time  should break down at 
very short distances of the order of the Planck length. Field theories on noncommutative spaces may
 play an important role in unraveling the properties of nature at
 the Planck scale. It has been shown that the noncommutative
geometry naturally appears in string theory with a non zero 
 antisymmetric B-field.

Beside the string theory arguments the noncommutative field
theories are very interesting on their own right. In a noncommutative
space-time the coordinate operators satisfy the relation
\begin{equation}
[{\hat x}^{\mu},{\hat x}^{\nu}]=i\theta^{\mu\nu},
\end{equation}
where $\hat x$ are the coordinate operators and $\theta^{\mu\nu}$
 is an antisymmetric tensor of dimension  (length)$^2$.
  In general  noncommutative version of a field theory is obtained by
 replacing the product of the fields appearing in the action by
the star products
\begin{equation}
(f \star g)(x)=\exp \left( \frac{i}{2} \theta^{\mu\nu}
\frac{\partial}{\partial x^\mu}\frac{\partial}{\partial
y^\nu}\right) f(x)g(y) \mid_{x=y},
\end{equation}
where $f$ and $g$ are two arbitrary functions that we assume to
be infinitely differentiable.\\
 In recent years there have been a
 lot of work devoted to the study of noncommutative field theory
  and  noncommutative quantum mechanics, and  possible experimental
  consequences of extensions of the standard formalism (see the
reviews [1]  and references therein). Apart from this there has been also a growing interest
 in  possible cosmological consequences
of space non-commutativity. Here we focus on Reissner-Nordstrom (Re-No) 
 black hole in noncommutative spaces.\\
In a recent work [2], the authors studied the Re-No black hole in
 noncommutative spaces. They argued that using commutation
 relations (1) and coordinate transformation
$x_{i}=\hat{x}_{i}+\frac{1}{2}\theta_{ij}\hat{p}_{j}$,
$p_{i}=\hat{p}_{i}$, where $p_{i}$ and $x_{i}$ satisfy the usual
 commutation relations of quantum mechanics,  the Re-No black hole
 can be extended to  noncommutative spaces. By a substitution
 of radial coordinate in terms of its noncommutative equivalent
 $r\rightarrow \hat{r}=\hat{x}_{i}\hat{x}_{i}$, the authors
 derived a line element for Re-No black hole in a noncommutative
 space and studied its thermodynamics.  The main problem regarding their line element is : It does not seem  to be       solution  of Einstein equations. Then the question arises that  what is the relevant equation,  and what are the        definition of energy and temperature for this new equation?. There seems to be  no modified Einstein
 equations in this case, so the physical relevance of the
 resulting line element is obscure. Another unclear point is that once
 $\hat{r}$ is written in terms of the matrix $\theta^{ij}$ and the
 conventional position operators $x_{i}$ and  momenta $p_{i}$,
 $ds^{2}$ is  far from what we mean by a line element.\\
 Another  important point is that the proposed line element (see
 section $4$ in [2])  exhibits, by the presence of the charge, a behaviour worse than  $\frac{1}{r^{4}}$, with an        inconsistent spherical symmetry breaking. And finally one more unconvincing result
 regarding this perturbative expansion (in $\theta$ parameter)
 approach is  that curvature singularities   continue to exist in spite of introducing a minimal length. Coordinate      noncommutativity implies the existence of a finite minimal length $\sqrt{\theta}$,
 below which  concept of ``distance'' becomes physically
 meaningless. This underlines the problem  to define the
 line element, namely the infinitesimal
 distance between two nearby points in Einstein gravity.\\
 In section $(2)$ we study the  Re-No black hole in noncommutative
 spaces that solve  the above mentioned inconsistencies. We begin by a brief  review  of   Re-No black hole in commutative spaces.\\
 A spherically symmetric solution of the coupled Einstein and Maxwell equations is that of Reissner and Nordstrom, 
 which represents a black hole with  mass $M$ and  charge $Q$.\\
 The metric of the  Re-No black hole is given by :\footnote{We
 have employed Gaussian units along with natural units, and set 
 Newton's constant  to unity.}
\begin{equation}
\label{4} ds^2= \left( 1 - \frac{2M}{r} + \frac{Q^2}{r^2} \right)
dt^2 - \frac{dr^2}{\left( 1 - \frac{2M}{r} + \frac{Q^2}{r^2}
\right)}- r^2 d \Omega^2.
\end{equation}
There are two apparent singularities at
\begin{equation}
\label{5} r_{\pm} = M \pm \sqrt{M^2 - Q^2},
\end{equation}
provided $M \geq Q$. Cosmic censorship dictates  this
 inequality, and hence there is an external event horizon at
 $r_{+}$. The other horizon $r_{-}$ is the internal Cauchy horizon.
 The limiting case when $Q=M$ and $r_{+}=r_{-}$ is referred to as
 the extremal case.\\

\section{Reissner-Nordstrom black hole in noncommutative spaces.}  

To analyze black holes in the framework of noncommutative spaces
one has to solve corresponding field equations. It is argued [3,4] that  it is not necessary to change the Einstein tensor part of the field equations, and that the noncommutative effects act only on the matter source. The underlying
 philosophy of this approach is to modify the distribution of point
 like sources in favor of smeared objects. This is in agreement
 with the conventional procedure for the regularization of UV
 divergences by introducing a cut off. Thus we conclude that in
 general relativity, the effect of noncommutativity can be taken
 into account by keeping the standard form of the Einstein tensor
 in the left-hand side (l.h.s) of the field equation and introducing a modified
 energy momentum tensor as a source in the right-hand side (r.h.s). This is exactly
 the gravitational analogue of the noncommutative  modification of quantum field
 theory [5]. For the  reasons mentioned in the previous section, we have developed an effective approach where
 noncommutativity is implemented only through a Gaussian
 de-localization of matter sources. In this way there  no problem arises in defining line element and Einstein           equations are kept
 unchanged.  We can summarize the approach as follows : a) in
 noncommutative geometry there cannot be point-like object,
 because there is no  physical distance  smaller than a minimal
 position uncertainty of the order of $\sqrt{\theta}$. b) this
 effect is implemented in space-time through matter
 de-localization, which by explicit calculations [5] turns out to be
 of Gaussian form. c) Space-time geometry is determined through
 Einstein's  equations with de-localized matter sources. d)
 de-localization of matter results in a regular, i.e. curvature
 singularity free, metric. This is exactly what is expected from
 the existence of a minimal length.\\
 The effect of smearing is mathematically implemented as a ``substitution rule" : position Dirac-delta function
  is replaced everywhere with a Gaussian distribution of minimal width $\sqrt{\theta}$. Inspired by this result, we
  choose the mass density of a static,
  spherically symmetric, smeared, particle-like gravitational source
  as [3,4]:
 \begin{equation}
 \rho_{\theta}(r)=\frac{M}{(4\pi\theta)^{\frac{3}{2}}}exp(\frac{-r^{2}}{4\theta}).
 \end{equation}

A particle of mass $M$, instead of being perfectly localized at a
 point is diffused throughout a region of     line size $\sqrt{\theta}$. This is due to the intrinsic uncertainty encoded in the coordinates commutator (1).\\

By solving the Einstein equations  with $\rho_{\theta}(r)$, as a
 matter source, we find the line element :
\begin{equation}
ds^{2}=-g_{00}dt^{2}+g^{-1}_{00}dr^{2}+r^{2}d\Omega^{2},
\end{equation}
where :
\begin{equation}
g_{00}=1-\frac{4M}{\sqrt{\pi}r}\gamma(\frac{3}{2},\frac{r^{2}}{4\theta})+\frac{Q^{2}}{\pi
r^{2}}\gamma^{2}(\frac{1}{2},\frac{r^{2}}{4\theta})-\frac{Q^{2}}{\pi
r\sqrt{2\theta}}\gamma(\frac{1}{2},\frac{r^{2}}{2\theta}),
\end{equation}
and
\begin{equation}
\gamma(\frac{a}{b},x)\equiv \int^{x}_{0}\frac{du}{u}u^{a/b}e^{-u},
\end{equation}
is the lower incomplete Gamma function.\\
 In the limit $\frac{r}{\sqrt{\theta}}\rightarrow\infty$, we get the classical Re-No metric i.e. the Re-No metric
  in commutative spaces. The line element (6) describes the geometry of a noncommutative Re-No black hole and gives
  us useful information about possible noncommutativity effects on the properties of this type of black hole. Using
  equation $g_{00}(r_{H})=0$, one can find the  event horizon(s):\\

$r_{\pm}=\frac{2M}{\sqrt{\pi}}\gamma(\frac{3}{2},\frac{r^{2}}{4\theta})+$

\begin{equation}
\frac{Q^{2}}{2\pi\sqrt{2\theta}}\gamma(\frac{1}{2},\frac{r^{2}}{2\theta})\pm
\frac{1}{2}\left[\left(\frac{4M}{\sqrt{\pi}}\gamma(\frac{3}{2},\frac{r^{2}}{4\theta})+\frac{Q^{2}}
{\pi\sqrt{2\theta}}\gamma(\frac{1}{2},\frac{r^{2}}{2\theta})\right)^{2}+\frac{4Q^{2}}{\pi
}\gamma^{2}
(\frac{1}{2},\frac{r^{2}}{4\theta})\right]^{\frac{1}{2}}.
\end{equation}

It is convenient to invert Eq.(9) and consider the black hole mass
 $M$ as a function of $r_{H}$ :

\begin{equation}
M=\frac{Q^{2}}{2\sqrt{2\pi\theta}}+\frac{1}{\gamma\left(\frac{1}{2},\frac{r_{H}^{2}}{4\theta}\right)}\left[\frac{\sqrt{\pi}}{4}r_{H}+\frac{Q^{2}}{4\sqrt{\pi}r_{H}}G(r_{H})\right],
\end{equation}\\
where
\begin{equation}
G(r)\equiv
\gamma^{2}\left(\frac{1}{2},\frac{r^{2}}{4\theta}\right)-\frac{r}{\sqrt{2\theta}}\gamma\left(\frac{1}{2},\frac{r^{2}}{2\theta}\right).
\end{equation}
 In the limit $\frac{r_{H}}{\sqrt{\theta}}<<1$, where one expects significant changes due to
 space non-commutativity, Eq.(10) leads to
 \begin{equation}
 M\rightarrow M_{0}\approx 0.5 \sqrt{\pi\theta}+0.2\frac{Q^{2}}{\sqrt{\pi\theta}},
 \end{equation}
 which is an interesting result. Noncommutativity implies a minimal non-zero mass that allows the existence
  of an event horizon. If the  black hole has an initial  mass  $M>M_{0}$, it can radiate until the value $M_{0}$ 
  is reached. At  this point the horizon has totally evaporated leaving behind a massive relic. Since black holes with
  mass $M<M_{0}$ do not exist there are three possibilities :\\
1.  For $M>M_{0}$ there is a black hole with regular metric in the origin.\\
2.  For $M=M_{0}$ the event horizon shrinks to zero.\\
3.  For $M<M_{0}$ there is no horizon.\\

 The reason why it does not end-up into a naked singularity is due to the finiteness of the curvature at the origin.
  Compared to the neutral black hole ($Q=0$) the effect of charge is to increase the minimal non-zero mass.

  The physical nature of the mass $M_{0}$ remnant is clearer  if we consider the black hole temperature as
  a function of  $r_{H}$. We have:\\

 $T_{H}\equiv\left(\frac{1}{4\pi}\frac{dg_{00}}{dr}\right)_{r=r_{H}}=$
\begin{equation}
\frac{1}{4\pi r_{H}}\left[1-N(\theta)-\frac{4Q^{2}}{\pi
r_{H}^{3}}\gamma^{2}(\frac{3}{2},\frac{r_{H}^{2}}
{4\theta})-\frac{Q^{2}}{\pi r_{H}^{3}}N(\theta)G(r_{H})\right],
\end{equation}
where :

\begin{equation}
N(\theta)=\frac{r_{H}^{3}exp(\frac{-r_{H}^{2}}{4\theta})}{4\theta^{\frac{3}{2}}
\gamma(\frac{3}{2},\frac{r_{H}^{2}}{4\theta})}.
\end{equation}

In the large  radius limit, i.e. $\frac{r_{H}}{\sqrt{\theta}}>>1$,
 one recovers the standard result for the Hawking temperature :
\begin{equation}
T_{H}=\frac{1}{4\pi r_{H}}-\frac{Q^{2}}{16\pi
r^{3}_{H}}=\frac{1}{4\pi
r_{H}}\left(1-\frac{Q^{2}}{4r^{2}_{H}}\right).
\end{equation}
On the other hand in the limit
$\frac{r_{H}}{\sqrt{\theta}}\rightarrow 0$, we have :
\begin{equation}
T_{H}\propto
\frac{r_{H}}{\pi\theta}\hspace{2.cm}\frac{r_{H}}{\sqrt{\theta}}\rightarrow
0.
\end{equation}
Eqs (15) and (16) are very interesting and have two
 important consequences. First, when the black hole completely evaporates it reaches  zero temperature  and there will 
  be no  horizon. Second, it reaches a maximum temperature while passing from the regime of  large radius to the
 regime of small radius. This is the same behaviour encountered in
 the noncommutative neutral case [3].
 The effect of charge is just to lower the maximum temperature, see Eq.(15).\\

\section{Specific heat, free energy and thermodynamic stability.}

 A black hole as a thermodynamic system is unstable if it has
 negative specific heat. We study the thermodynamic stability of
 noncommutative Re-No black hole by evaluating its
 specific heat and free energy.\\
 We know that the entropy is proportional to the area of event
horizon :
\begin{equation}
S=\frac{A}{4}=\pi r_{H}^{2}.
\end{equation}
Using the first law of thermodynamics $dE=T dS+\Phi dq$, where
 $\Phi$ is the electrostatic  potential, we obtain the
 following expression for the energy :
\begin{equation}
E= M_{0}+2\pi \int_{r_{0}}^{r_{H}}r_{H}'' T(r_{H}'') dr_{H}''+
\int_{r_{0}}^{r_{H}} \Phi(r_{H}'') dq(r_{H}''),
\end{equation}
where $M_{0}$ is the minimal mass below which no black hole can
 be formed and $r_{0}$ is the minimal horizon, see Eq.(10).\\
 In order to check the stability of the noncommutative Re-No black
 hole we evaluate the heat capacity :
\begin{equation}
C_{v}=\frac{\partial E(r_{H})}{\partial T(r_{H})}=(\frac{\partial
E(r_{H}) }{\partial r_{H}})(\frac{1}{\frac{\partial
T(r_{H})}{\partial r_{H}}}).
\end{equation}
$E$ increases  monotically as $r$ increases, but as mentioned
 earlier $T$ has a maximum at $r=r_{max}$. Below(above) $r_{max}$, T  is  a  monotonically  increasing (decreasing) 
 function of $r$. The heat capacity is positive for
 $r_{0}<r_{H}<r_{max}$ and negative for $r_{H}>r_{max}$. Thus  the
 black hole is stable if $r_{0}<r_{H}<r_{max}$,  and is unstable if
 $r_{H}>r_{max}$.\\
 The free energy of the noncommutative Re-No black hole is given
 by :
\begin{equation}
F=E(r_{H})-T(r_{H}) S(r_{H}).
\end{equation}
By evaluating $F$, and using the fact that the black hole is
 stable(unstable) when the free energy has a local minimum(maximum), we again  see that for $r_{H}< r_{max}$
 the black hole is stable while it is unstable if $r_{H}>r_{max}$.\\

\section{The upper bound on the noncommutativity parameter $\theta$.}

Using Eq.(12) and the extremal condition $M_{ext}=Q$, one can find
 an upper bound on the noncommutativity parameter $\theta$. We
 have :
\begin{equation}
0.5 \sqrt{\pi\theta}+0.2\frac{Q^{2}}{\sqrt{\pi\theta}} \geq
M_{ext}=Q,
\end{equation}

which gives the following upper bound for the noncommutativity
 parameter $\theta$ :
\begin{equation}
\theta \le 0.02 Q^{2}.
\end{equation}
It is also interesting to discuss our expectation about the lower bound for the parameter $\theta$. As mentioned         earlier passing from the regime of large radius to the regime of small radius, Eqs (15) and (16) imply the existence    of a maximum temperature. The role of charge is to lower the maximum temperature.\\
 In commutative case one expects relevant back-reaction effects during the terminal stage of evaporation because of      huge increase of temperature . As it has been shown, the role of noncommutativity is to cool down the black hole in     the final stage. As a consequence [4], there is a suppression of quantum back-reaction since the black hole emits less  and less energy. But  back-reaction may be important during the maximum temperature phase. In order to estimate its     importance in this region, we consider  the thermal energy $E= T_{H}$  and the total mass $M$. In order to have        
 significant back-reaction effect $T^{Max.}_{H}$ should be of the same order of magnitude as $M$. For the neutral case   i.e. $Q=0$, from Eqs (10) and (13)  we have $M\cong 2.4 \sqrt{\theta}M^{2}_{Pl.}$  and  $T_{H}^{max.}=1.5\times            10^{-2}/\sqrt{\theta}$, so  we shall obtain the following estimation :
\begin{equation}
\sqrt\theta\approx 10^{-1}\ell_{Pl.} = 10^{-34} cm.
\end{equation}
For the case of charged black holes,  the role of charge is to lower the maximum temperature. Therefore we obtain  even    smaller values  for the noncommutativity parameter $\theta$. Expected values of $\sqrt\theta$ are well above the        Planck length $\ell_{Pl.}$, so (23) (and the smaller values for the case of charged black holes)  indicate that          back-reaction effects are suppressed  if  $\sqrt\theta\approx 10\ell_{Pl.}$ (or  even  $\sqrt\theta > 10\ell_{Pl.}$ for the case of charged black holes).  For this reason we can safely use unmodified form of the
 metric (6) during all the evaporation process. So, we can safely consider $\sqrt\theta\geq 10^{-33} cm$.\\

\section{Discussion.}

In this section we discuss two important issues. First, since the concept of a black hole is inherently
 coordinate-independent, and since the restriction to space-space
 noncommutativity implies the choice of  a  specific  space-time
 slicing, it is not obvious that the inferred modifications of
 black hole properties are coordinate independent features. Then the question is how we can justify the general          covariance of the results.  The modifications
 occurring at the level of energy momentum tensor (EMT) do not
 modify its tensorial properties. In other words, the
 noncommutativity provides  a fluid type EMT instead  of the
 conventional EMT generating the Schwarzschild solution, which  is
 wrongly considered a vacuum solution [6,8]. We only
 need to solve the Einstein equations  plugging this new EMT in the
 same way as consider a cosmological fluid in  the Robertson-Walker space-time. Of course these coordinates
  coincide with the Schwarzschild  spherical coordinates
 as  can be seen from  the solution slightly away  from the
 origin. Therefore there is no  problem with  coordinate
 independence once the derivation is
 tensorially consistent.\\

Second, how we can implement noncommutativity  by  changing only the matter part of  Einstein equation 
     and leaving the left hand side of the equation intact. 
  One of  the  main differences  between noncommutative and
  commutative  theories  stems from the fact
  that  in a noncommutative space the coordinates   operators have no common
  position eigenvectors  due to equation (1). 
 It has been  known  since the seminal work of Glauber in quantum optics  
 [7], that there exist coherent states that  are eigenstates of
 annihilation operator. As already stated, the reason behind use
 of coherent states is that there are no coordinate eigenstates
 for NC coordinates and no coordinate representation can be
 defined. Therefore, ordinary wave functions (in quantum mechanics) or fields
 defined over points (in Quantum field theory) can not be defined anymore. Coherent
 states are the closest to the sharp coordinate states that one
 can define for NC coordinates in the sense that they are
 minimal-uncertainty states and enable us to define mean values of coordinate
 operators. Coherent states, properly defined as
 eigenstates of ladder operators built from noncommutative
 coordinate operators only, are the closest to the sharp
 coordinate states, which we can define for non-commutative
 coordinates. This means that  coordinate coherent states are the
 minimal uncertainty states over the noncommutative manifold and
 let us calculate the aforementioned mean values [8]. This implies that  the matter  field is also
 modified, since now it has to be written in terms of ``mean
 coordinates'', eventhough  ``formally''  it is left unchanged.\\

\section{Conclusions}

In conclusion we have studied the Re-No black hole in
 noncommutative spaces. We have found the Re-No metric and Hawking
 temperature in noncommutative spaces that  reproduce exactly
 ordinary Re-No solution at large
 distances($\frac{r}{\sqrt{\theta}}\rightarrow \infty$).  We have
 shown that like the neutral case there is a minimal non-zero  mass  $
M_{0}\approx 0.5
\sqrt{\pi\theta}+0.2\frac{Q^{2}}{\sqrt{\pi\theta}}$ to which a
 black hole can decay through radiation. The effect of charge $Q$ 
 is  to increase this  minimal  mass. The reason why it does
 not end-up into a naked singularity is due to the finiteness of
 the curvature at the origin. From thermodynamics point of view,
 the same kind of regularization takes place eliminating the
 divergent behaviour of  Hawking temperature. As a consequence, 
 there is a maximum temperature that the black hole can reach
 before cooling down to absolute zero. The effect of charge $Q$ is to lower this  maximum temperature.\\
 We  have also  found an upper  bound  for  the noncommutativity parameter $\theta$.\\

\section{Acknowledgment.}

 I would like to thank R. Allahverdi (UNM) for proofreading of this manuscript.\\

\section{References.}
1. R. Szabo, Class. Quant. Grav. 23(2006)R199-R242 and 
   Phys. Rept. 378(2003)207.\\
2. K. Nozari, B. Fazlpour, ActaPhys.Polon.B39(2008)1363.\\
3. P. Nicolini, J. Phys. A38 (2005) L631-L638.\\
4.  P. Nicolini et.al., Phys. Lett. B632 (2006)547, S. Ansoldi, et.al., Phys. Lett. B645 (2007)261.\\
5. A. Smailagic, E. Spalluci, J. Phys. A36(2003)L517, J.Phys. A37 (2004)1; Erratum-ibid. A37 (2004) 7169.,
J. Phys. A36(2003)L467.\\
6. H. Balasin and H. Nachbagauer, Class. Quant. Grav. 10, (1993)2271.\\
7. R.J. Glauber Phys. Rev.131(1963)2766.\\
8. P. Nicolini, arXiv:0807.1939.\\

\end{document}